\documentclass{vgtc}                          

\ifpdf
\pdfoutput=1\relax                   
\usepackage{graphicx}                
\DeclareGraphicsExtensions{.pdf,.png,.jpg,.jpeg} 
\else
\usepackage{graphicx}                
\DeclareGraphicsExtensions{.eps}     
\fi%

\graphicspath{{figures/}{pictures/}{images/}{./}} 

\usepackage{microtype}                 
\PassOptionsToPackage{warn}{textcomp}  
\usepackage{textcomp}                  
\usepackage{mathptmx}                  
\usepackage{times}                     
\usepackage{cite}                      
\usepackage{tabu}                      
\usepackage{booktabs}                  
\usepackage[babel]{csquotes}
\usepackage{wasysym}
\usepackage{amssymb}
\usepackage{enumitem}

\usepackage{todonotes}

\usepackage{array, colortbl, makecell, rotating} 

\newcommand{\symVeryLow}{{\fontsize{4}{6}\selectfont $\blacksquare~\square~\square~\square~\square$}}
\newcommand{\symLow}{{\fontsize{4}{6}\selectfont $\blacksquare~\blacksquare~\square~\square~\square$}}
\newcommand{\symMedium}{{\fontsize{4}{6}\selectfont $\blacksquare~\blacksquare~\blacksquare~\square~\square$}}
\newcommand{\symHigh}{{\fontsize{4}{6}\selectfont $\blacksquare~\blacksquare~\blacksquare~\blacksquare~\square$}}
\newcommand{\symVeryHigh}{{\fontsize{4}{6}\selectfont $\blacksquare~\blacksquare~\blacksquare~\blacksquare~\blacksquare$}}

\newcommand{\symYes}{{\Large \CIRCLE}}
\newcommand{\symNo}{{\Large \Circle}}
\newcommand{\symPartial}{{\Large \LEFTcircle}}

\DeclareRobustCommand{\inlinenlpsymbol}[1]{%
	\begingroup\normalfont
	\raisebox{-.2\height}{\includegraphics[height=0.9\fontcharht\font`\D]{figures/symbols/#1}}
	\endgroup
}

\newcommand{\symCaseStudy}{\hspace{-0.1em}\inlinenlpsymbol{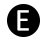}\hspace{-0.1em}}
\newcommand{\symComparison}{\hspace{-0.1em}\inlinenlpsymbol{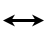}\hspace{-0.2em}}
\newcommand{\symInterview}{\hspace{-0.1em}\inlinenlpsymbol{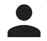}\hspace{-0.1em}}

\onlineid{0}

\vgtccategory{Evaluation}

\vgtcinsertpkg

\title{Towards a Survey on Static and Dynamic Hypergraph Visualizations}

\author{Maximilian T.\ Fischer~\thanks{max.fischer@uni-konstanz.de} %
	\and Alexander Frings~\thanks{alexander.frings@uni-konstanz.de} %
	\and Daniel A.\ Keim~\thanks{keim@uni-konstanz.de} %
	\and Daniel Seebacher~\thanks{daniel.seebacher@uni-konstanz.de}%
}
\affiliation{\scriptsize University of Konstanz, Germany}

\abstract{
	Leveraging hypergraph structures to model advanced processes has gained much attention over the last few years in many areas, ranging from protein-interaction in computational biology to image retrieval using machine learning.
	Hypergraph models can provide a more accurate representation of the underlying processes while reducing the overall number of links compared to regular representations.
	However, interactive visualization methods for hypergraphs and hypergraph-based models have rarely been explored or systematically analyzed.
	This paper reviews the existing research landscape for hypergraph and hypergraph model visualizations and assesses the currently employed techniques.
	We provide an overview and a categorization of proposed approaches, focusing on performance, scalability, interaction support, successful evaluation, and the ability to represent different underlying data structures, including a recent demand for a temporal representation of interaction networks and their improvements beyond graph-based methods.
	Lastly, we discuss the strengths and weaknesses of the approaches and give an insight into the future challenges arising in this emerging research field.
} 

\keywords{Hypergraphs, hypergraph model, temporal, visualization, visual analytics, survey}

\begin{document}
	
	
	\firstsection{Introduction}
	\maketitle
	The modeling of complex network structures has risen to prominence over the last few years~\cite{Wang.NetworkSecurityGraphModel.2007, Vehlow.Survey.VisGroupGraphs.2017}.
	While the usage of graph models has dominated the field, recently, a trend has emerged for models to leverage hypergraphs instead of graphs to more accurately represent the underlying problem structure~\cite{Klamt.HypergraphCellularNetw.2009}.
	As hypergraphs generalize graphs by extending edges to connect any number of vertices, groups and many-to-many relationships can be captured more efficiently and in more detail, while neither compromising nor leaving out data points and connections~\cite{Valdivia.ParallelAggOrdHypergraphVis.2019}.
	Hypergraph modeling lifts the boundary of limited interconnections while preserving the ability to encode other attributes like directed or weighted edges.
	
	Therefore, hypergraphs have been employed in many areas ranging from network security~\cite{Wang.NetworkSecurityGraphModel.2007} to studying protein-protein interactions in computational biology~\cite{Przulj.PPIGraphModeling.2011} or feature selection in medical diagnosis~\cite{Somu.HypergraphFeatureSelection.2016} to path-signaling in cell interaction models~\cite{Ritz.SignalingHypergraphs.2014}.
	Furthermore, neural networks based on hypergraph structures have been shown~\cite{Feng.Hypergraph.2019, Jiang.DynamicHypergraphNeuralNetw.2019} to yield increased performance on some tasks such as network classification and object recognition through their more complex and high-order correlations.
	A primary benefit is the reduced number of interconnections necessary to represent strongly interrelated networks.
	For example, 
	human communication and social media data can be represented more effectively~\cite{Onnela.HumCommGraph.2007, Heintz.BeyondGraphs.2014, Amato.InfluenceSMNHypergraphs.2017} by directly representing internal group structures and leveraging superset-to-subset combinations~\cite{Heintz.BeyondGraphs.2014}.
	
	While modeling problems as hypergraphs can provide benefits, the visualization of the hypergraphs themselves or even the hypergraph models are not trivial tasks~\cite{Heintz.BeyondGraphs.2014}.
	Mathematically, an undirected hypergraph $H = (V, E)$ is defined as an ordered pair, where $V = \{v_1,.., v_n\}$ represents the $n$ vertices (hypernodes) and subsets of these vertices form the multi-set $E = \{e_1,.., e_m\}$, constituting the $m$ distinct hyperedges~\cite{Berge.Hypergraphs.1984}.
	We do not explicitly differentiate between classes of hyperedges, i.e. we do not consider heterogeneous hypergraphs separately. 
	Traditionally, the visualization of such hypergraphs uses Venn diagrams or Euler diagrams.
	As part of the field of graph drawing, their traditional representation has been described in detail~\cite{Johnson.VennDrawing.1987, Makinen.DrawingHypergraph.1990, Bertault.DrawingHypergraphsSubset.2001}.
	However, as these traditional representations often use color to distinguish between hyperedges, their scalability is severely limited.
	While hypergraph visualizations can share some characteristics with multivariate graphs~\cite{Ward.InteractiveDataVisualization.2015}, the strategies are not always suitable for hypergraphs, which need a more efficient way of depiction~\cite{Heintz.BeyondGraphs.2014}.
	The problem of hypergraph visualization has even become more complex by the recent interest in the evolution of such relations.
	These temporal (or dynamic) hypergraphs additionally encode time for direct comparisons of network states in different stages and require new visualization approaches for an effective representation.
	However, more advanced hypergraph visualizations are relatively new and not systematically explored, lacking detailed comparisons between approaches so far.
	
	To address these issues, we present a survey of hypergraph visualizations, making the
	following contributions:
	\begin{itemize}[topsep=0.5em]
		\setlength{\itemsep}{0pt}
		\setlength{\parskip}{0.125em}
		\setlength{\parsep}{0pt}
		\item a systematic literature review of existing approaches for static and dynamic hypergraph (model) visualization.
		\item a methodology for comparison criteria between hypergraphs, critically assessing the different approaches.
	\end{itemize}
	
	This survey aims to provide insights into the existing research landscape of generic hypergraph visualizations and identify promising research gaps for future work.
	
	\section{Related Work}
	Work on more advanced hypergraph visualizations has only recently gained traction.
	Consequently, most surveys cover hypergraph-visualizations only as a sub-part -- if at all -- and rarely provide an extensive overview of the existing methodologies with respective advantages and drawbacks.
	Therefore, to compile a list of existing approaches, we also consider more generic surveys on set- and (regular) graph visualizations~\cite{Landesberger.STAR.VALargeGraphs.2011, Alsallakh.Survey.SetVisualization.2016, Beck.Survey.DynamicGraphsVis.2017, Vehlow.Survey.VisGroupGraphs.2017, Chen.Survey.VisAssocRelGraphs.2019, McGee.SurveyMultilayerNetwVis.2019} which focus on relational aspects, as some aspects are comparable.
	
	Alsallakh et al. provide a comprehensive overview~\cite{Alsallakh.Survey.SetVisualization.2016} of set visualizations, which can be regarded as a specific representation of hypergraphs.
	Many of these approaches are different or enhanced variants of Euler and Venn diagrams like BubbleSets~\cite{Collins.BubbleSets.2009}, but it also contains several different node-link and matrix-based approaches.
	However, not all apply to generic hypergraphs, as they are sometimes very domain-specific.
	In this survey, we only cover generic hypergraph visualizations.
	
	Vehlow et al. analyzed group visualizations using different regular graph representations in their survey~\cite{Vehlow.Survey.VisGroupGraphs.2017}.
	They propose a taxonomy for visualization techniques separated into visual node attributes, juxtaposed, superimposed, and embedded visual styles.
	This survey forms the groundwork for contrasting group structures and helps in categorizing static hypergraph approaches, covering several basic representations but does not include more recent approaches.
	
	When shifting the focus on dynamic graphs, one of the latest surveys~\cite{Beck.Survey.DynamicGraphsVis.2017} on dynamic graph visualizations was compiled by Beck et al. in 2017, focusing on group structures in regular graphs and their relationship over time.
	They differentiate between animated node-link diagrams and timeline structures to convey the dynamic relationships visually.
	The authors also state that approaches utilizing timelines are becoming more common in the literature.
	
	For a more set theory-focused approach, the survey~\cite{Chen.Survey.VisAssocRelGraphs.2019} by Chen et al. focuses on exploring association relationships in graphs.
	The authors propose a pipeline for visual analysis of associated data and summarized many different graph representations for large relationship data sets.
	Furthermore, they discussed the advantages and disadvantages of some visualization types, including interaction combined with simplification methods.
	As part of their work, they list two visualization methods for hypergraphs, one being the radial layout by Kerren et al.~\cite{Kerren.RadialVisHypergraphs.2013}, the other being a fixed-node layout presented by Xia et al.~\cite{Xia.HypergraphBone.2011}.
	However, this survey seems incomplete and somewhat outdated regarding hypergraph visualizations, as it did not cover some available approaches.
	
	Also, as new hypergraph visualization methods have been proposed since these works have been released and the lack of an extensive survey for hypergraphs, it becomes apparent that there is a gap for an overview of such techniques specifically.
	Although many proposed pipelines and taxonomies can also be applied to hypergraph models, they do not provide in-depth comparison grounds for them and their dynamic counterparts.
	For example, Vehlow et al.~\cite{Vehlow.Survey.VisGroupGraphs.2017} covered most basic representations, but recent hybrids~\cite{Streeb.VAHypergraphPrediciton.2019, Valdivia.ParallelAggOrdHypergraphVis.2019, Fischer.HyperMatrix.2020} are partly incompatible with their differentiation criteria.
	
	
	
	Recently Valdivia et al. begin to tackle the research gap by
	proposing PAOHvis~\cite{Valdivia.ParallelAggOrdHypergraphVis.2019}, thereby providing the \enquote{first [...] highly readable representation of dynamic hypergraphs}.
	It features a timeline-like view and linearly orders the intervals to directly compare the hyperedges and the whole structure side-by-side.
	This approach, by design, is especially suited for a medium-sized (less than 100) number of relatively small hyperedges, favoring comparability \emph{over time} instead of \emph{at one time}.
	For dynamic hypergraphs, a hybrid visualization~\cite{Streeb.VAHypergraphPrediciton.2019} was proposed by Streeb one year earlier.
	They use arrow-glyphs for time-frames encoding the network changes.
	
	The idea was improved and then generalized to arbitrary dynamic hypergraph \emph{models} by Fischer et al.~\cite{Fischer.HyperMatrix.2020}.
	Hyper-Matrix consists of a geometric deep learning model architecture and presents a novel multi-level, matrix-based hypergraph visualization.
	
	\section{Methodology}
	To analyze state-of-the-art approaches on hypergraph visualizations, we conducted a survey of relevant literature before collecting further approaches via cross-references.
	We target the subset of visualization literature that applies to hypergraphs and hypergraph models, and those techniques that focus on representing hypergraph structures and how they are reflected and realized in the visualization.
	
	\subsection{Source and Selection Methodology}
\begin{figure}
		\centering%
		\includegraphics[width=\linewidth]{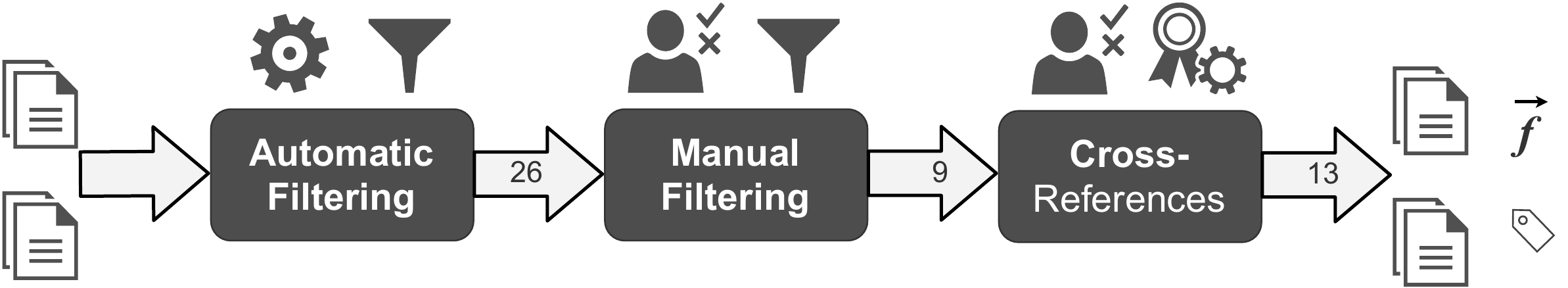}
		\caption{\textbf{The paper collection process} consists of three main steps: (1) Automated filtering, (2) manual filtering, and (3) cross-referencing.}
		\label{fig:paper_selection_process}
	\end{figure}

	Orienting ourselves on semi-automated structured literature surveys, such as the one of Sacha et al.~\cite{Sacha.VisualInteractionDR.2017}, we initialize the paper collection process using a keyword-based search for \enquote{hypergraph}, and common variations like \enquote{hyper-graph, hyper graph, ...}. To further focus our survey, we consider only approaches from the following high-quality journals and conferences since 2000:
	\begin{itemize}[topsep=0.5em]
		\setlength{\itemsep}{0pt}
		\setlength{\parskip}{0.125em}
		\setlength{\parsep}{0pt}
		\item IEEE Transactions of Visualization and Computer Graphics (\textbf{TVCG}, incl. \textbf{IEEE VIS} proceedings)
		\item Computer Graphics Forum (CGF, incl. \textbf{EuroVis} and EuroVA)
	\end{itemize}
	For the actual paper selection methodology, we follow a three-step approach (see also Figure~\ref{fig:paper_selection_process}).
	Automated filtering resulted in 5 (TVCG) and 21 (CGF) approaches; manual filtering reduced this first to 3 and 10, while in a second iteration after careful considerations, to 3 and 6, i.e., nine approaches. Using cross-references and survey data, we identified an additional four relevant approaches, leading to 14 approaches in total, which we compare using the following comparison framework.
	
	\subsection{Comparison Framework}
	We present seven comparison criteria covering some aspects common to graphs and some favoring aspects specifically adapted to hypergraphs.
	We curated this selection based on an adapted combination of criteria taken from existing surveys~\cite{Vehlow.Survey.VisGroupGraphs.2017, Beck.Survey.DynamicGraphsVis.2017} as well as aspects discussed in the literature~\cite{Valdivia.ParallelAggOrdHypergraphVis.2019, Fischer.HyperMatrix.2020}, and complemented them by additional distinguishing criteria.
	

	
	\textbf{Representation Method}
	The first criterion provides a rough separation into visualization types, thereby deciding on the fundamental technique. The existing approaches can be classed as either \emph{node-link} diagrams, \emph{timeline-based} techniques, or \emph{matrix-based} approaches.
	
	
	\textbf{Scalability}
	The supported data-set sizes and, accordingly, the scalability of the approaches (nodes and hyperedges) present a significant factor.
	Diving the capabilities of existing approaches in five distinguishable groups, we distinguish between very low scalability (\symVeryLow, less than ten nodes or hyperedges), low scalability (\symLow, between 10 and 50), medium scalability (\symMedium, between 50 to 200), high scalability (\symHigh, between 200 and 1000), and very high scalability (\symVeryHigh, more than 1000).
	Especially node-link-diagrams with too many connections are prone to clutter.
	Many basic representations are only applicable to few vertices and hyperedges, whereas different techniques perform better.
	
	\textbf{Static vs. Dynamic}
	This criterion is concerned with the support of dynamic (i.e., temporal) data (\symYes) in contrast to static data (\symNo).
	Analyzing trends requires different time steps to be comparable in the visualization.
	The methods range from a simple timeline to a direct comparison of single components by highlighting different stages.
	Nevertheless, approaches supporting dynamic hypergraphs are still limited and only recently began to emerge.
	
	\textbf{Interactivity}
	This criterion describes the support of interactivity (\symYes) by visualizations compared to a static, not modifiable (\symNo) visualization.
	It describes the ability to modify the model or the resulting representation for analysis purposes.
	Besides highlighting, sorting or filtering can enable the analysis of larger graphs.
	For example, even Venn diagrams would benefit from interactivity to reduce ambiguities. 
	Other visualization methods may depend on a multi-level structure to represent the entire hypergraph coherently.
	
	\textbf{Tasks \& Use Cases}
	This criterion describes if the approaches present tasks in the application domain and their specific uses.
	For example, temporal networks of cell interactions may use comparisons of the sub-graph as small multiples to outline its development.
	
	\textbf{Evaluation}
	The last comparison criterion concerns evaluations, case studies, and user studies conducted on the different visualization methods.
	Only a few approaches provide no evaluation at all (\symNo).
	However, some evaluations are performed as case studies (\inlinenlpsymbol{icon_case_study}~), for example, by comparing the scalability to other representation types like parallel coordinate plots to demonstrate hypergraph model advantages. In contrast, others conduct both a full user study (\inlinenlpsymbol{icon_person}) with participants, often domain experts, concerned with performing analytical tasks, but rarely a quantitative study.

	\section{Literature Survey}
	For the following survey, we group the individual techniques based on the representation method and, secondly, their publication date.
	
	\subsection{Node-Link-based Approaches}
	
	\textbf{Bubble Sets}
	Collins et al.~\cite{Collins.BubbleSets.2009} describe hulls with contouring (marching squares) encompassing connected elements. 
	Interaction is provided through movement and the addition or deletion of nodes.
	
	

	\textbf{Software Artifact Hypergraph Visualization}
	Kapec et al. presented a visualization of software functions in a visual programming environment, using hypergraphs as a basis~\cite{Kapec.SoftwareArtifactsHypergraphs.2010}.
	Source code and calling sequences are modeled as hypergraph-connections using a force-directed 3D layout with spheres and directed links, while color is used to distinguish callers and callees.  
	
	
	
	\textbf{EGAN Hypergraph Visualization}
	Paquette et al.~\cite{Paquette.HypergraphVisualization.2011} propose an extensible visualization method for Exploratory Gene Association Networks (EGAN) to sort and classify gene lists.
	It uses a node-link-based approach but adds a separate association meta-node that connects to each related entity, imitating the concept of hyperedges, thereby improving scalability compared to color-based methods and being similar to node-link representations.

	\textbf{Kelp Diagram}
	Dinkla et al.~\cite{Dinkla.KelpDiagrams.2012} extend on the concept of colored-hulls and bubble-maps by using methods like overlapping lines, node color, or varying link sizes.
	It is primarily suited for static and fixed, geospatial applications, not providing interactions, and limited in scalability.
	A successor, KelpFusion~\cite{Meulemans.KelpFusion.2013}, improves performance while reducing clutter, combining hull and linear set representation.

	\textbf{Radial Representation}
	Kerren et al.~\cite{Kerren.RadialVisHypergraphs.2013} present a different layout where hyperedges are circular dotted lines around centered nodes, thereby eliminating overlaps and slightly increasing the scalability.
	The visualization offers rich interaction support from highlighting groupings, filtering, and link modification and was evaluated in a small user study.
	The authors present possible extensions, for example, labeling, graph comparison, or an extended study.
	
	\textbf{Visual Analysis of Set Relations in a Graph}
	Xu et al.~\cite{Xu.SetRelations.2013} present a glyph-based overlay in node-link diagrams to study the effect between shared set relations and node distances in a graph.
	However, scalability is very limited and interaction concepts are only discussed as future work.
	
	\textbf{Extra-Node Representation}
	Ouvrard et al.~\cite{Ouvrard.HypergraphModeling.2017} present an improved method for node-link conversion of hypergraphs by adding artificial extra-nodes which combine and merge multiple edge connections, retaining more information and inducing less clutter, which is shown qualitatively and quantitatively.
	
	\textbf{SimpleHypergraphs.jl}
	SimpleHypergraph.jil is a julia library by Antelmi et al.~\cite{Antelmi.SimpleHypergraphsjl.2020}, offering established interactive hypergraph visualizations using a combination of D3 and Python.
	The visualization approaches are similar to EGAN~\cite{Paquette.HypergraphVisualization.2011}, but representing hyperedges as sub-graphs, but also support a convex hull approach similar to a Venn diagram, and support some limited interactivity.
	

	\textbf{MetroSets}
	The MetroSets approach by Jacobsen et al.~\cite{Jacobsen.MetroSets.2021} represents geospatial hypergraph data, and its primary application is the visualization of train networks.
	The drawbacks of the model are its enforced octolinearity and unavoidable crossings leading to clutter.
	Interaction support ranges from highlighting to filtering different sets, but not for node repositioning.
	As one of a few approaches, it supports temporal data mapped to the x-axis, representing hyperedges at different time-steps.
	
	
	\subsection{Timeline-based Approach}
	
	\textbf{TimeSets}
	Nguyen et al.~\cite{Nguyen.TimeSets.2016} leverage a stacked timeline view with hyperedges encoded via color, similar to KelpFusion.
	It offers different interaction techniques, a strong case as well as user study, but suffers from poor scalability without aggregating.
	
	\textbf{PAOHvis}
	Valdivia et al. introduce PAOHvis~\cite{Valdivia.ParallelAggOrdHypergraphVis.2019}, an ordered timeline layout to represent hyperedges.
	It is based on earlier work~\cite{Valdivia.DynamicHypergraphs.2018}, follows a grid timeline pattern and nodes on the y-axis.
	Hyperedges are represented through vertical lines connecting multiple nodes, where drip dots can mitigate overflowing.
	The layout optimizes space usage and allows for parallel vertical hyperedges to split into multiple groups, offering rich filtering and interaction controls like timeflow comparison.

	\textbf{Set Streams}
	Agarwal et al.~\cite{Agarwal.SetStreams.2020} leverage branching and merging streams in a timeline view to represent dynamic set membership.
	Different set operations can be applied through queries, and linked lists offer more detailed information, while the scalability extends to several hundred elements.
	
	\subsection{Matrix-based Approach}
	\textbf{Visual Analytic Framework for Dynamic Hypergraphs}
	Streeb et al.~\cite{Streeb.VAHypergraphPrediciton.2019} propose a prototype for temporal hypergraph analysis by leveraging a glyph-based matrix view.
	It supports temporal data through time glyphs but may require interaction to retrieve information, and its scalability is limited.
	
	\textbf{Hyper-Matrix}
	The latest visualization method was proposed by Fischer et al.~\cite{Fischer.HyperMatrix.2020} and leverages a six-level matrix-based representation for visualizing dynamic hypergraph structures.
	A geometric deep learning model interfaces with a hypergraph model, and the interface supports multiple filtering, grouping, and interaction mechanism, including matrix-reordering, hierarchical grouping, and the direct integration of domain knowledge to influence the underlying machine learning model.

	\begin{table*}[ht!]
		\centering  
		
		\renewcommand{\arraystretch}{0.95}
		\setlength{\tabcolsep}{4pt}
		\begin{tabular}{r|c|c|c|c|c|c|c|c|c|c|c|c|c|c|} 
			
			& \multicolumn{9}{c|}{\textbf{Node-Link} } 
			& \multicolumn{3}{c|}{\textbf{Timeline} } & \multicolumn{2}{c|}{\textbf{Matrix} } \\
			

			& \multicolumn{1}{c}{\begin{turn}{90} \parbox[t]{1.35cm}{\small BubbleSets \newline\cite{Collins.BubbleSets.2009}} \end{turn}}     
			& \multicolumn{1}{c}{\begin{turn}{90} \parbox[t]{1.35cm}{\small Kapec \newline\cite{Kapec.SoftwareArtifactsHypergraphs.2010}} \end{turn}} 
			& \multicolumn{1}{c}{\begin{turn}{90} \parbox[t]{1.35cm}{\small Paquette \newline\cite{Paquette.HypergraphVisualization.2011}}  \end{turn}}
			& \multicolumn{1}{c}{\begin{turn}{90} \parbox[t]{1.35cm}{\small Kelp D. \newline\cite{Dinkla.KelpDiagrams.2012}} \end{turn}}
			& \multicolumn{1}{c}{\begin{turn}{90} \parbox[t]{1.35cm}{\small Kerren \newline\cite{Kerren.RadialVisHypergraphs.2013}}  \end{turn}} 
			& \multicolumn{1}{c}{\begin{turn}{90} \parbox[t]{1.35cm}{\small Xu \newline\cite{Xu.SetRelations.2013}}\end{turn}} 
			& \multicolumn{1}{c}{\begin{turn}{90} \parbox[t]{1.35cm}{\small Ouvrard \newline\cite{Ouvrard.HypergraphModeling.2017}}\end{turn}} 
			& \multicolumn{1}{c}{\begin{turn}{90} \parbox[t]{1.35cm}{\small Antelmi \newline\cite{Antelmi.SimpleHypergraphsjl.2020}} \end{turn}} 
			& \multicolumn{1}{c|}{\begin{turn}{90} \parbox[t]{1.35cm}{\small MetroSets \newline\cite{Jacobsen.MetroSets.2021}} \end{turn}} 
			& \multicolumn{1}{c}{\begin{turn}{90} \parbox[t]{1.35cm}{\small TimeSets \newline\cite{Nguyen.TimeSets.2016}} \end{turn}}
			& \multicolumn{1}{c}{\begin{turn}{90} \parbox[t]{1.35cm}{\small PAOHvis \newline \cite{Valdivia.ParallelAggOrdHypergraphVis.2019}}\end{turn}}
			& \multicolumn{1}{c|}{\begin{turn}{90} \parbox[t]{1.35cm}{\small Set Streams \newline \cite{Agarwal.SetStreams.2020}}\end{turn}}
			& \multicolumn{1}{c}{\begin{turn}{90} \parbox[t]{1.35cm}{\small Streeb\newline \cite{Streeb.VAHypergraphPrediciton.2019}}\end{turn}} 
			& \multicolumn{1}{c|}{\begin{turn}{90} \parbox[t]{1.35cm}{\small HyperMatrix \newline \cite{Fischer.HyperMatrix.2020} }\end{turn}}
			\\ \toprule
			
			
			Scalability & 
			\symVeryLow &\symLow & \symVeryLow & \symVeryLow & \symLow & \symMedium
			& \symHigh & \symLow & \symLow 
			& \symLow & \symMedium & \symHigh & 
			\symMedium & \symHigh \\ \midrule
			
			Domain & 
			{\small Geo/Netw.} & {\small Softw.} &  {\small Bio.} & {\small Geo/Netw.} & {\small Generic} & 	{\small Netw.}
			& {\small Netw.} & {\small Netw.} & {\small Geo} 
			& {\small Events} & {\small History} & {\small Netw.} & 
			{\small Comms} & {\small Comms} \\ \midrule
			
			Dynamic Support & 
			\symNo &\symNo & \symNo  & \symNo & \symPartial & \symNo
			& \symNo & \symNo & \symYes 
			& \symYes & \symYes & \symYes & 
			\symYes & \symYes \\ \midrule
			
			Interactivity & 
			\symNo &\symYes & \symPartial  & \symNo & \symPartial & \symNo
			& \symNo & \symPartial & \symNo
			& \symYes & \symYes & \symYes & 
			\symYes & \symYes \\ \midrule
			
			{ Tasks \& Use Cases} &
			\symYes & \symYes & \symPartial  & \symPartial & \symPartial & \symYes
			& \symNo & \symPartial & \symPartial
			& \symYes & \symYes & \symYes & 
			\symPartial & \symYes \\ \midrule
			
			Evaluation & 
			\symCaseStudy & \scalebox{0.6}{\symNo} & \symCaseStudy &  \symComparison & \symInterview & \symCaseStudy &
			\symCaseStudy\symComparison & \symCaseStudy & \symCaseStudy\symComparison
			& \symCaseStudy\symComparison\symInterview  & \symCaseStudy\symInterview & \symCaseStudy\symInterview & 
			\scalebox{0.6}{\symNo} & \symCaseStudy\symComparison\symInterview \\ \bottomrule
		\end{tabular}
		
		\vspace{0.25cm}
		\centering  
		\renewcommand{\arraystretch}{1.2} 
		{ \small
			\begin{tabular}{p{1.72cm}p{1.2cm}p{1.5cm}p{1.15cm}p{1.7cm}|p{0.9cm}p{0.8cm}p{1.1cm}|p{0.7cm}p{1.5cm}p{0.9cm}} 
				\raisebox{.2\height}{\symVeryHigh}~Very High &
				\raisebox{.2\height}{\symHigh}~High &
				\raisebox{.2\height}{\symMedium}~Medium &
				\raisebox{.2\height}{\symLow}~Low &
				\raisebox{.2\height}{\symVeryLow}~Very Low   &

				\scalebox{0.8}{\symYes}~Present &
				\scalebox{0.8}{\symPartial}~Partly & 
				\scalebox{0.8}{\symNo}~Missing &
				
				\scalebox{1.6}{\raisebox{.11\height}{\symCaseStudy}}~Case  &
				\scalebox{1.6}{\raisebox{.11\height}{\symComparison}}~Comparison & 
				\scalebox{1.6}{\raisebox{.11\height}{\symInterview}}~User   \\
			\end{tabular}
		}
		\vspace{0.15cm}
		\caption{Overview and comparison of hypergraph visualization techniques, grouped according to primary visualization technique.}
		\label{Table:Comparison}
	\end{table*}

	
	\subsection{Further Methods}
	During our initial search for related work (searching for \emph{hypergraph} on Google Scholar), we discovered some related approaches in different disciplines.
	In the following, we shortly reference approaches where the focus is a task-dependent problem related to a \emph{specific} hypergraph representation in a different domain and not focused on visualization research, and therefore not full-filling the inclusion criteria of this survey for a generic and generalizable approach.
	
	Early works by Jin et al.~\cite{Jin.OverlappingMatrixPatternHypergraph.2008} present combinations of hypergraph and matrix visualizations through hyperrectangles.
	Arafat et al.~\cite{Arafat.HypergraphForceDirected.2017} adapt force-directed placement for increased scalability of hull drawing, while Cooper et al.~\cite{Cooper.MultilevelTopicDepHypergraph.2019} using hypergraphs for topic dependency sketching through varying thickness and colored lines for the association.
	Cromar et al.~\cite{Cromar.Hyperscape.2015} visualize protein-protein interactions using a combination of Venn and node-link diagrams, while Liu et al.~\cite{Liu.UncertaintyAwareMicroblogRetrieval.2016} uses clustered glyphs layouts for uncertainty modeling.

	\section{Comparison and Discussion}
	
	In this survey, we studied currently proposed visualization techniques for hypergraphs and hypergraph models.
	Table~\ref{Table:Comparison} provides an overview of the different approaches and highlights commonalities as well as differences.
	The presented approaches utilize different interactive features to extend their techniques.
	In general, node-link-based approaches can only support edge counts ranging from few (Venn diagram) up to several dozens, depending on the technique. In comparison to this, timeline- and matrix-based approaches can support medium- and large-sized data-sets with several hundred nodes and hyperedges.
	Additionally, there is a clear evolution, as dynamic support and advanced interactivity are featured more prominently in more recent techniques.
	Interestingly, user studies have also become more common over time.
	Finally, the aim mainly revolves around improving performance compared to non-hypergraph visualizations.
	
	Most hypergraph visualizations are based on node-link diagrams (or variants thereof) and use regular graph drawing with special nodes or hulls.
	The work by Kapec~\cite{Kapec.SoftwareArtifactsHypergraphs.2010} sticks out as an early technique providing an extended, three-dimensional, and scalable model compared to its contemporaries.
	However, all styles use different colors or shapes on their sub-graphs as visual variables to encode node affiliations.
	Consequently, some methods have difficulties presenting an accurate picture of the structure as they rely on overlapping projections or might lose information about hyperedges in the drawing process.
	The interactive adaptations boil down to manipulating positions or viewpoints of the model.
	The main drawback is the scalability, as many methods become  illegible quickly.
	
	The more advanced node-link-based visualizations offer more accurate, powerful, and scalable methods, encoding the hypergraph (models) using an adaption of preexisting techniques.
	For example, all of them use regular edges.
	However, they should be interpreted as forming hyperedges through the nodes, sometimes with additional visuals for improved clarity.
	The exception is the Kelp-diagram approach~\cite{Dinkla.KelpDiagrams.2012}, as it specializes in fixed point visualizations useful for marking geospatial information.
	However, this heavily limits its applicability and scalability when compared to the other two approaches.
	For example, the radial representation~\cite{Kerren.RadialVisHypergraphs.2013} can visualize more hyperedges side-by-side while the MetroSets~\cite{Jacobsen.MetroSets.2021} deal with overlap when excessively extended.
	They provide basic-level interactions like highlighting, necessary for interpreting larger data sets.
	Still, only the MetroSets~\cite{Jacobsen.MetroSets.2021} features labels applied to both nodes and hyperedges.
	Further, they are also the only technique that incorporates support for dynamic data, while radial representations may only partly support it.
	Nevertheless, the amount of information this style can display is increased compared to the earlier approaches.
	
	The most common tasks are the search for nodes in the visualization~\cite{Kapec.SoftwareArtifactsHypergraphs.2010, Paquette.HypergraphVisualization.2011, Dinkla.KelpDiagrams.2012, Kerren.RadialVisHypergraphs.2013, Xu.SetRelations.2013, Nguyen.TimeSets.2016, Streeb.VAHypergraphPrediciton.2019, Valdivia.ParallelAggOrdHypergraphVis.2019, Fischer.HyperMatrix.2020, Jacobsen.MetroSets.2021} and the determination of communities/memberships and connected components~\cite{Paquette.HypergraphVisualization.2011, Dinkla.KelpDiagrams.2012, Xu.SetRelations.2013, Nguyen.TimeSets.2016, Streeb.VAHypergraphPrediciton.2019, Valdivia.ParallelAggOrdHypergraphVis.2019, Antelmi.SimpleHypergraphsjl.2020, Agarwal.SetStreams.2020, Fischer.HyperMatrix.2020,  Jacobsen.MetroSets.2021}, sometimes to determine their size~\cite{Nguyen.TimeSets.2016}.
	Other common techniques are membership queries~\cite{Collins.BubbleSets.2009, Paquette.HypergraphVisualization.2011, Xu.SetRelations.2013, Nguyen.TimeSets.2016, Streeb.VAHypergraphPrediciton.2019, Valdivia.ParallelAggOrdHypergraphVis.2019, Fischer.HyperMatrix.2020} and the tracing of connections~\cite{Collins.BubbleSets.2009, Kerren.RadialVisHypergraphs.2013, Nguyen.TimeSets.2016, Valdivia.ParallelAggOrdHypergraphVis.2019, Agarwal.SetStreams.2020, Fischer.HyperMatrix.2020, Jacobsen.MetroSets.2021}. The latter is sometimes extended to determine implicit shares and overlap~\cite{Xu.SetRelations.2013} and random walks~\cite{Antelmi.SimpleHypergraphsjl.2020}.
	Less frequent tasks are filtering based on external factors~\cite{Kerren.RadialVisHypergraphs.2013, Agarwal.SetStreams.2020, Fischer.HyperMatrix.2020}, evaluating the readability~\cite{Nguyen.TimeSets.2016}, classification~\cite{Agarwal.SetStreams.2020}, and change detection in models~\cite{Fischer.HyperMatrix.2020}.

	
	The matrix-based visualizations feature more scalable techniques to display large correlated data-sets.
	They additionally provide extensive interactions to higher degrees, needed to work with the visualizations in realistic scenarios.
	The PAOHvis approach~\cite{Valdivia.ParallelAggOrdHypergraphVis.2019} uses a column-wise timeline representation of hyperedges while the other two methods compare the nodes and their affiliations directly in the cells.
	To achieve this, they use the nodes as coordinates while different layers of time or tags represent the underlying data.
	Still, Streeb et al.'s technique~\cite{Streeb.VAHypergraphPrediciton.2019} is the only one to render a classic node-link-based model as part of their visualization.
	The latest technique, Hyper-Matrix~\cite{Fischer.HyperMatrix.2020}, is equally easy to handle but focuses more on connection development for predicting future affiliations and extends the concept to a generic blueprint for hypergraph model visualization through a new architecture and novel visualization technique.
	
	\section{Future Directions}
	Even with recent advances, opportunities for future work remain.
	The node-link-based approaches provide a very limited \textbf{scalability}, and even timeline~\cite{Agarwal.SetStreams.2020} and matrix~\cite{Fischer.HyperMatrix.2020} representations do not support more than around thousand entities.
	Solving these issues requires novel ideas like extra-nodes~\cite{Ouvrard.HypergraphModeling.2017}, aggregating and subsetting, as well as dense, domain-specific representations.
	Support for \textbf{dynamic} hypergraphs is offered only by six approaches and is often optimized for specific use cases. 
	Specifically, the support for dynamic hypergraphs \emph{with many time steps} is very limited.
	One could, for example, explore how novel concepts in dynamic networks visualizations beyond animations~\cite{Cakmak.MultiscaleSnapshots.2020} are applicable to hypergraphs.
	Further, there is no established \textbf{benchmark dataset} for hypergraph visualizations, no established \textbf{performance metrics}, and only a limited discussion~\cite{Valdivia.ParallelAggOrdHypergraphVis.2019, Fischer.HyperMatrix.2020} on specific \textbf{tasks}.
	
	%
	
	\section{Conclusion}
	In the last decade, problem modeling through hypergraphs has gained much attention.
	While some visualization methods have been proposed, there is a noticeable and increasing gap between applied hypergraph research and their visualization.
	By surveying the existing approaches for 
	hypergraph (model) visualizations, we aim to structure the research space.
	We first establish the need for such a survey by analyzing existing literature before defining a reproducible paper selection process.
	Then, we systematically structure comparison criteria 
	before presenting the techniques individually.
	We discuss the particularities of each technique individually, before classifying it, and finally discuss in detail the observations in context.
	We find that many representations do not capture the full potential of hypergraphs and are limited in scalability, interactivity, or the support of dynamic hypergraphs.
	The three most promising and generic techniques, PAOHvis~\cite{Valdivia.ParallelAggOrdHypergraphVis.2019}, Set Streams~\cite{Agarwal.SetStreams.2020}, and Hyper-Matrix~\cite{Fischer.HyperMatrix.2020} are also the most recent ones, but all approaches need further refinement.
	This supplements our observations and highlights the specific opportunities for future work we have identified.
	
	By filling this gap with an overview of hypergraph visualization methods, we aim to provide researchers with a standard reference,
	promote areas for future work, and set the baseline for a more in-depth survey on hypergraph visualizations.
	
	\acknowledgments{\noindent The authors acknowledge the financial support by the Federal Ministry of Education and Research of Germany (BMBF) in the framework of PEGASUS under the program "Forschung für die zivile Sicherheit 2018 - 2023" and its announcement "Zivile Sicherheit - Schutz vor organisierter Kriminalität II". } 

	\pagebreak
	\bibliographystyle{abbrv-doi}
	\bibliography{bibliography}

\end{document}